\documentclass{emulateapj}

\slugcomment{Draft Version \today}
\shorttitle{Mapping Overionization in W49B}
\shortauthors{LOPEZ ET AL.}

\newcommand{\ltsima}{$\; \buildrel < \over \sim \;$}
\newcommand{\simlt}{\lower.5ex\hbox{\ltsima}}

\newcommand{\ls}{{_<\atop^{\sim}}}

\newcommand{\gs}{{_>\atop^{\sim}}}

\newcommand{\Te}{$kT_{\rm e}$}
\newcommand{\Tej}{$kT_{\rm ej}$}
\newcommand{\Tz}{$kT_{\rm z}$}

\begin{document}

\title{Unraveling the Origin of Overionized Plasma in the Galactic Supernova Remnant W49B}

\author{Laura A. Lopez\altaffilmark{1,5,6}, Sarah Pearson\altaffilmark{2}, Enrico Ramirez-Ruiz\altaffilmark{3}, Daniel Castro\altaffilmark{1}, Hiroya Yamaguchi\altaffilmark{4}, Patrick O. Slane\altaffilmark{4}, Randall K. Smith\altaffilmark{4}}
\altaffiltext{1}{MIT-Kavli Institute for Astrophysics and Space Research, 77 Massachusetts Avenue, 37-664H, Cambridge MA 02139, USA}
\altaffiltext{2}{Dark Cosmology Centre, Niels Bohr Institute, University of Copenhagen, Juliane Maries Vej 30, 2100, Copenhagen, Denmark}
\altaffiltext{3}{Department of Astronomy and Astrophysics, University of California Santa Cruz, 1156 High Street, Santa Cruz, CA 95060, USA}
\altaffiltext{4}{Harvard-Smithsonian Center for Astrophysics, 60 Garden St., Cambridge, MA 02138, USA}
\altaffiltext{5}{NASA Einstein Fellow}
\altaffiltext{6}{Pappalardo Fellow in Physics}

\email{lopez@space.mit.edu}

\begin{abstract}

Recent observations have shown several supernova remnants (SNRs) have overionized plasmas, those where ions are stripped of more electrons than they would be if in equilibrium with the electron temperature. Rapid electron cooling is necessary to produce this situation, yet the physical origin of that cooling remains uncertain. To assess the cooling scenario responsible for overionization, in this paper, we identify and map the overionized plasma in the Galactic SNR W49B based on a 220 ks {\it Chandra} Advanced CCD Imaging Spectrometer (ACIS) observation. We performed a spatially-resolved spectroscopic analysis, measuring the electron temperature by modeling the continuum and comparing it to the temperature given by the flux ratio of the He-like and H-like lines of sulfur and of argon. Using these results, we find that W49B is overionized in the west, with a gradient of increasing overionization from east to west. As the ejecta expansion is impeded by molecular material in the east but not in the west, our overionization maps suggest the dominant cooling mechanism is adiabatic expansion of the hot plasma. 

\end{abstract}


\section{Introduction}

Young supernova remnants (SNRs) offer the means to study stellar explosions and their interactions with the interstellar medium (ISM). The kinetic energy of these explosions is dissipated in collisionless shocks that heat the ejecta and ISM gas to X-ray emitting temperatures. In young SNRs, these shocks create an X-ray emitting, ionizing plasma that slowly reaches collisional ionization equilibrium (CIE). However, X-ray observations with {\it ASCA} first revealed possible evidence for ``overionized'' remnants, where the electron temperature \Te\ derived from the continuum is systematically lower than the temperature \Tz\ given by the line ratios \citep{kawasaki02,kawasaki05}. The presence of radiative recombination continuum (RRC) features, discovered recently with {\it Suzaku} observations of various SNRs, provided conclusive evidence of overionization\citep{yamaguchi09,ozawa09,ohnishi11,sawada12,uchida12,yamauchi13}.  
 
In a collisional plasma (as in SNRs), this overionized plasma is one signature of rapid electron cooling, and the physical origin of this cooling in SNRs remains a topic of debate. One scenario where cooling can occur is thermal conduction, in which the hot ejecta in the SNR interior may cool by exchanging heat efficiently with the exterior material (e.g., \citealt{cox99,shelton99}). Alternatively, another possible cooling mechanism is adiabatic expansion, where the SN blast wave expands through a dense circumstellar medium (CSM) into a rarefied ISM (e.g., \citealt{itoh89,moriya12,shimizu12}). \cite{kawasaki02,kawasaki05} interpreted their {\it ASCA} results as evidence of cooling via thermal conduction, whereas \cite{yamaguchi09} demonstrated the timescale for this scenario is too slow to explain the recombining plasma observed with {\it Suzaku}. 

Observationally, one may distinguish which cooling scenario is responsible by mapping ionization state across SNRs. For example, if overionized plasma is concentrated where it is in contact with colder material, it would suggest cooling via thermal conduction. Conversely, if overionized plasma is less prevalent where ejecta are confined, the plasma likely cooled through adiabatic expansion. 

\begin{figure*}
\includegraphics[width=\textwidth]{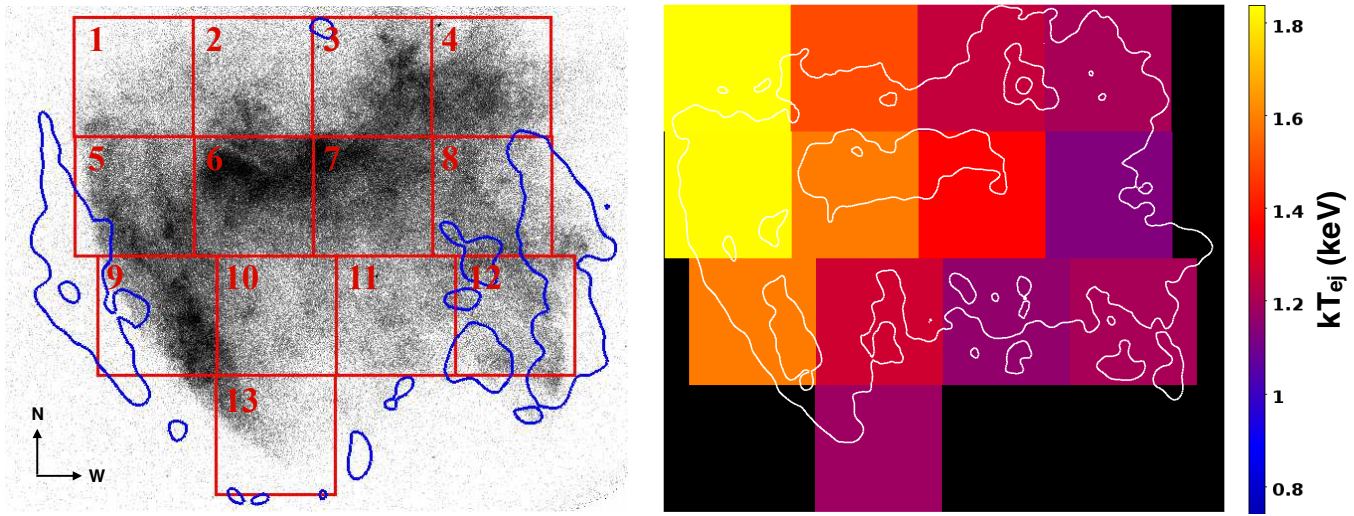}
\caption{Left: 13 regions (in red) of area 59$\arcsec \times 59\arcsec$ analyzed to map overionization based on He-like to H-like line flux ratios. Regions are over plotted on the full-band (0.5--8.0 keV) {\it Chandra} ACIS X-ray image. Blue contours show the location of warm H$_{2}$ (at 2.12 $\mu$m; from \citealt{keohane07}). In the east, the ejecta are impacting molecular material, while the ejecta in the west are expanding into lower-density ISM (note that the molecular material seen in the west is likely in the foreground: \citealt{lacey01,lopez12}). Right: Hot plasma electron temperature of the ejecta $kT_{\rm ej}$ estimated by modeling the continuum emission across W49B, with X-ray full-band contours overplotted in white.}
\label{fig:kT_e}
\end{figure*}

W49B (G43.3$-$0.2) is one of the first SNRs where signatures of a recombining plasma were discovered. Using integrated {\it ASCA} spectra, \cite{kawasaki05} measured the intensity ratios of He-like to H-like lines of Ar and Ca, and they found $kT_{\rm z} \sim$ 2.5 keV $\gtrsim kT_{\rm ej} \sim$ 1.8 keV, where $kT_{\rm ej}$ is the ejecta electron temperature. \cite{miceli06} performed similar analyses on the global {\it XMM-Newton} spectra, and they found no evidence for overionization. However, they cautioned that the temperature distribution in W49B is not uniform and noted that a spatially-resolved search for overionization may yield different results. Further evidence in favor of recombining plasma in W49B was discovered by \cite{ozawa09} using {\it Suzaku} observations: they found an RRC feature from He-like Fe at 8.830 keV and derived $kT_{\rm z} \sim$ 2.7 keV. More recently, \cite{miceli10} attempted to localize spatially the recombining plasma using the archival {\it XMM-Newton} data by mapping the hardness ratio of the pn count rate in the 8.3--12 keV band to that in the 4.4--6.2 keV band. They found that the hardness ratio is elevated in the center and west of W49B, which these authors attribute to the Fe RRC. However, the limited counts in the {\it XMM-Newton} observations precluded a spatially-resolved spectral modeling of these features, and thus the elevated hardness ratio could be from other sources or processes (e.g., high background, harder bremsstrahlung, non-thermal emission). Therefore, it is critical to confirm the results of \cite{miceli10} via another means to constrain the cooling scenario responsible for overionization. Additionally, given that the Fe as a distinct spatial distribution from the other metals in W49B \citep{lopez09a,lopez11}, it is worth while to explore whether lighter elements are also overionized in the same locations. 

We note that to date, W49B is the only ejecta-dominated SNR and possibly the youngest source where these features have been observed. The age estimate for W49B is uncertain, ranging from $\sim$1,000 years \citep{pye84} up to $\sim$6,000 years \citep{smith85}. Other SNRs identified to have recombining plasmas are more dynamically mature and have ages $\sim$4,000--20,000 years (see $\S$10.3 of the review by \citealt{vink12}). Thus, the overionized state of W49B, a dynamically young SNR, suggests the physical origin of the recombining plasma can occur in the initial phases of SNR evolution. 

In this paper, we identify and map the overionized plasma in W49B in order to constrain the physical origin of the rapid cooling. For this analysis, we employ the recent 220 ks {\it Chandra}  X-ray Observatory Advanced CCD Imaging Spectrometer (ACIS) observation of W49B \citep{lopez12}. Although {\it Chandra} is dominated by background above $\sim$8 keV and thus is not capable of discerning the Fe RRC, we can identify the overionized plasma another way. In particular, we measure the ejecta electron temperature \Tej\ (from the bremsstrahlung continuum) and \Tz\ (from line flux ratios) across many regions to assess how the ionization state varies across the SNR. In Section~\ref{sec:data}, we detail the observations, the procedure we employ to estimate the temperatures, and the results. In Section~\ref{sec:discussion}, we discuss the implications regarding the role of the two cooling mechanisms in the SNR evolution and we consider alternate interpretations of the results. In Section~\ref{sec:summary}, we present a brief summary of this work and outline future observations that could further elucidate the physical origin of rapid cooling in W49B. 

\section{Mapping Overionization in W49B with Chandra} \label{sec:data}

W49B was observed for 220 ks with {\it Chandra} ACIS on 18--22 August 2011 with the backside-illuminated S3 chip in the Timed-Exposure Faint Mode (ObsIDs 13440 and 13441). Reprocessed data were downloaded from the {\it Chandra} archive, and analyses were performed using the {\it Chandra} Interactive Analysis of Observations ({\sc ciao}) Version 4.3. \cite{lopez12} reported recently a spatially-resolved spectroscopic analysis on these data; in the central 2.9\arcmin\, a total of $\sim$10$^{6}$ net counts (after background subtraction) were recorded in the 0.5--8.0 keV band during the observation. The integrated X-ray spectrum of W49B (see Figure~2 of \citealt{lopez12}) includes prominent He-like and H-like lines of silicon (Si {\sc xiii} and Si {\sc xiv}), sulfur (S {\sc xv} and S {\sc xvi}), argon (Ar {\sc xvii} and Ar {\sc xviii}), and calcium (Ca {\sc xix} and \hbox{Ca {\sc xx}}). 

For the analyses, we extracted spectra from 13 regions of area 59$\arcsec$ $\times$ 59$\arcsec$, as shown in Figure~\ref{fig:kT_e}. These regions each have $\gs$25000 net full-band counts, with $\gs$5000 and $\gs$2500 net counts in the S and Ar bands respectively, sufficient to fit these features successfully. We found that smaller regions did not give us enough counts to distinguish the Ar {\sc xviii} line (the weakest line we include in our analysis) above the continuum. We modeled spectra using XSPEC Version 12.7.0 \citep{arnaud96}, and we adopt the solar abundance values of \cite{asplund09}. The choice of solar abundance values affects the best-fit absorbing column $N_{\rm H}$, since X-rays are attenuated by ISM metals (see \citealt{wilms00} for a detailed discussion). 

\begin{deluxetable*}{lcccccccc} 
\tabletypesize{\footnotesize}
\tablecolumns{9} 
\tablecaption{Best-Fit Temperatures and Line Fluxes \label{table:lines}}
\tablewidth{0pt}
\tablehead{\colhead{Region\tablenotemark{a}} & \colhead{$kT_{\rm ISM}$} & \colhead{$kT_{\rm ej}$} & \colhead{S {\sc xv}} & \colhead{S {\sc xvi}} & \colhead{$kT_{\rm z, S}$\tablenotemark{c}} & \colhead{Ar {\sc xvii}} & \colhead{Ar {\sc xviii}} & \colhead{$kT_{\rm z, Ar}$\tablenotemark{c}} \\
 \colhead{} & \colhead{(keV)} & \colhead{(keV)} & \colhead{Flux\tablenotemark{b}} & \colhead{Flux\tablenotemark{b}} & \colhead{(keV)} & \colhead{Flux\tablenotemark{b}} & \colhead{Flux\tablenotemark{b}} & \colhead{(keV)}  }
\startdata
\hline \\
1 & 0.25$\pm$0.03 & 1.84$\pm$0.07 & 0.90 & 0.53 & 1.67 & 0.22 & 0.08 & 2.07 \\
2 & 0.24$\pm$0.01 & 1.50$\pm$0.03 & 1.12 & 0.63 & 1.65 & 0.26 & 0.09 & 2.05 \\
3 & 0.20$\pm$0.01 & 1.25$\pm$0.02 & 2.83 & 1.37 & 1.58 & 0.46 & 0.11 & 1.86 \\
4 & 0.17$\pm$0.01 & 1.20$\pm$0.02 & 2.65 & 1.37 & 1.61 & 0.49 & 0.12 & 1.89 \\
5 & 0.29$\pm$0.01 & 1.83$\pm$0.04 & 2.57 & 1.46 & 1.66 & 0.60 & 0.23 & 2.10 \\
6 & 0.21$\pm$0.01 & 1.59$\pm$0.02 & 5.81 & 3.60 & 1.70 & 1.22 & 0.44 & 2.07 \\
7 & 0.18$\pm$0.01 & 1.36$\pm$0.02 & 4.55 & 2.48 & 1.64 & 0.87 & 0.27 & 2.01 \\
8 & 0.16$\pm$0.02 & 1.13$\pm$0.01 & 3.07 & 1.29 & 1.50 & 0.45 & 0.14 & 1.99 \\
9 & 0.26$\pm$0.01 & 1.59$\pm$0.03 & 2.95 & 1.43 & 1.58 & 0.62 & 0.18 & 1.96 \\
10 & 0.23$\pm$0.01 & 1.27$\pm$0.03 & 2.04 & 0.85 & 1.49 & 0.37 & 0.11 & 1.97 \\
11 & 0.20$\pm$0.03 & 1.16$\pm$0.02 & 1.78 & 0.74 & 1.49 & 0.31 & 0.08 & 1.86 \\
12 & 0.14$\pm$0.01 & 1.20$\pm$0.02 & 1.61 & 0.74 & 1.55 & 0.32 & 0.10 & 2.00 \\
13 & 0.28$\pm$0.01 & 1.18$\pm$0.05 & 0.95 & 0.22 & 1.26 & 0.13 & 0.01 & 1.51 \\
\enddata
\tablenotetext{a}{Regions are numbered sequentially, as shown in Figure~\ref{fig:kT_e}.}  
\tablenotetext{b}{Fluxes are in units of 10$^{-4}$ ph cm$^{-2}$ s$^{-1}$}
\tablenotetext{c}{Temperature derived from the sulfur line ratios $kT_{\rm z, S}$ and from the argon line ratios $kT_{\rm z, Ar}$.}
\end{deluxetable*}

We obtain \Tej\ and \Tz\ of the ejecta using the following procedure. First, to determine $N_{\rm H}$ toward each region, we modeled the spectra as two absorbed CIE plasmas using the XSPEC components {\it phabs} and {\it vapec} v2.0.2 \citep{smith01,foster12}: one cooler, ISM component with solar abundances and one hotter, ejecta component with super-solar abundances of 3--10$\times$ the solar values of Si, S, Ar, Ca, and Fe by number \citep{hwang00,lopez09a}. We have selected this two-component model as previous X-ray studies of W49B have demonstrated that a solar abundance ISM plasma plus a ejecta plasma with super-solar abundances are necessary to adequately fit the spectrum \citep{hwang00,miceli06,lopez09a}. In our fits, $N_{\rm H}$, the cool and the hot component temperatures ($kT_{\rm ISM}$ and $kT_{\rm ej}$, respectively), and the abundances of elements with prominent emission lines (Si, S, Ar, Ca, and Fe) were allowed to vary freely. We found best-fit $N_{\rm H}$ of the 13 regions ranging 7.0--8.9$\times10^{22}$ cm$^{-2}$, consistent with the values derived in \cite{lopez12}. 

Previous X-ray studies of W49B with {\it ASCA} and {\it Chandra} \citep{hwang00,lopez09a} demonstrated that the Si {\sc xiii} emission may arise from the cool ISM component, while the other lines are predominantly from the ejecta component. As we want to measure the overionization of the ejecta, we should consider lines that originate from the hotter plasma. Based on the CIE spectral fits of the 13 regions performed above, we find that the fraction of the flux from the hot ejecta component in the Si {\sc xiii}, S {\sc xv}, and Ar {\sc xvii} lines is 55\%, 93\%, and 99\%, respectively. Therefore, in our estimates of the ionization temperature $kT_{\rm z}$ below, we consider the He-like and H-like lines of S and Ar. We did not include Ca as we determined that for the plasma temperatures of W49B, the systematic errors in the estimate of \Tz\ from the flux ratio of Ca {\sc xix} to Ca {\sc xx} were too large (as discussed below). We have divided the SNR into several regions and used the procedure outlined below to estimate $kT_{\rm ej}$ and $kT_{\rm z}$ at each location. 
 
We remodeled the spectra of each region phenomenologically, using two absorbed continuum components (one for the cool, ISM plasma and one for the hot, super-solar plasma) plus Gaussian functions to account for the line emission. To characterize the continuum of the two plasmas, we employed the AtomDB NoLine model\footnote{See http://www.atomdb.org/noline.php}, which removes the line emission from an {\it apec} plasma. The NoLine model includes three processes in modeling the continuum: bremsstrahlung, RRCs, and two photon emission from low-lying H- and He-like levels. Furthermore, we added Gaussian functions for fifteen emission lines with the following initial guesses for centroid energy: Mg {\sc xi} ($\approx$1.45 keV), Si {\sc xiii} ($\approx$1.85 keV), Si {\sc xiv} ($\approx$2.00 keV), Si K$\beta$ ($\approx$2.19 keV), S {\sc xv} ($\approx$2.45 keV), S {\sc xvi} ($\approx$2.60 keV), S K$\beta$ ($\approx$2.88 keV), Ar {\sc xvii} ($\approx$3.11 keV), Ar {\sc xviii} ($\approx$3.31 keV), Ar K$\beta$ ($\approx$3.69 keV), Ca {\sc xix} ($\approx$3.89 keV), Ca {\sc xx} ($\approx$4.13 keV), Ca K$\beta$ ($\approx$4.58 keV), Fe {\sc xxv} ($\approx$6.64 keV), and Fe K$\beta$ ($\approx$6.97 keV). To reduce the number of free parameters, we froze the $N_{\rm H}$ to the values derived from the two CIE plasma fits described above. We performed an initial fit by freezing the line centroid energies while letting the two plasma electron temperatures ($kT_{\rm ej}$ and $kT_{\rm ISM}$), plasma normalizations, and line fluxes vary. Subsequently, we thawed the line centroid energies and refit the spectra.  

\begin{figure*}
\includegraphics[width=\textwidth]{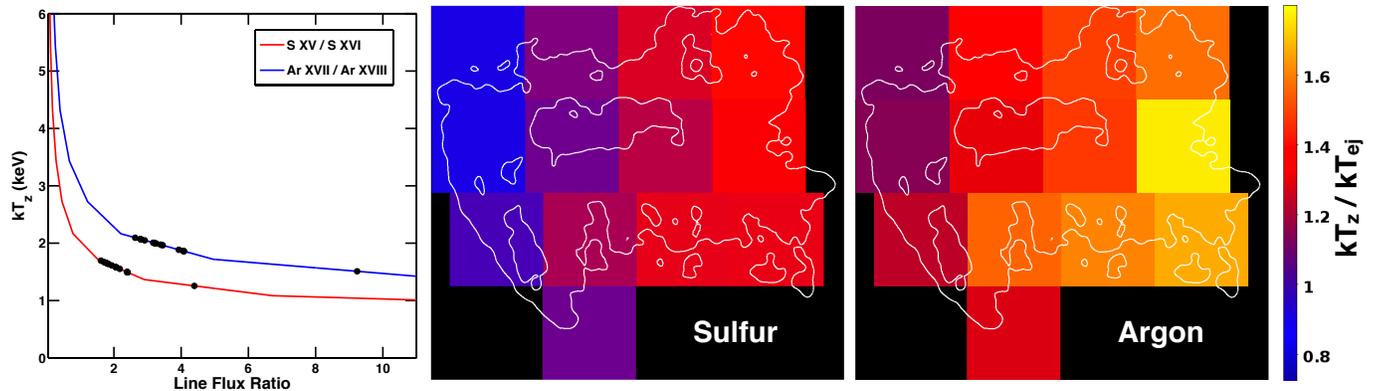}
\caption{Left panel: The line flux ratios (black data points) of He-like to H-like ions of S and Ar from regions across W49B. Solid lines are the conversion of these line ratios to ionization temperatures \Tz\ using current atomic data \citep{foster12}. Other panels: Maps of overionization across W49B. Middle panel: the ratio of ionization temperature \Tz\  (estimated using S {\sc xv} and S {\sc xvi} fluxes) and the electron temperature \Tej. Right panel: Maps of \Tz/\Tej\ using the flux ratio between Ar {\sc xvii} and Ar {\sc xviii}. Regions with  \Tz/\Tej $>$ 1 are overionized.}
\label{fig:overionization}
\end{figure*} 

Using this procedure, we obtained the two plasma electron temperatures (the hot ejecta temperature, $kT_{\rm ej}$, and the cool ISM temperature, $kT_{\rm ISM}$) and the line fluxes for the 13 regions, listed in Table~\ref{table:lines}. We found best-fit temperatures of the cool component ranging from $kT_{\rm ISM} \approx 0.14\pm$0.01 keV to $kT_{\rm ISM} \approx 0.29\pm$0.01 keV across the SNR. The best-fit ejecta temperatures also vary (see the map of $kT_{\rm ej}$ in the right panel of Figure~\ref{fig:kT_e}), with $kT_{\rm ej} \approx$1.13$\pm$0.01 keV up to $kT_{\rm ej} \approx$1.84$\pm$0.07 keV, with 90\% confidence errors. Our results suggest there is a temperature gradient, with rising $kT_{\rm ISM}$ and \Tej\ toward the east of the SNR. The expansion is thought to be impeded in the east by dense molecular material there \citep{keohane07}. The values we derive for $kT_{\rm ISM}$ and  \Tej\ in many regions of W49B are slightly lower than that estimated by \cite{kawasaki05}, who found $kT_{\rm ISM} = 0.24^{+0.04}_{-0.02}$ keV and \Te = 1.70$^{+0.02}_{-0.04}$ keV, from the integrated {\it ASCA} spectrum. We note that these authors assumed the continuum purely consisted of bremsstrahlung emission, whereas the NoLine model we have employed accounts for RRCs and two photon emission as well. The inclusion of these processes reduces the measured \Tej\ and also affects the derived \Tz\ from the line fluxes. Furthermore, our estimates of \Tej\ are lower than those estimated for the hot plasma component by \cite{miceli06} and \cite{lopez09a}, who find \Tej = 1.77--3.0 keV and \Tej = 1.80--3.68 keV, respectively. The reason is that the CIE fits performed by these authors use both the line ratios and the bremsstrahlung continuum to estimate \Tej. Thus, the fact that the derived \Tej\ in this previous work is greater than when \Tej\ is modeled from bremsstrahlung alone proves the overionized state of the plasma. 

We measured \Tz\ for the 13 regions based on the line flux ratios of the He-like and H-like lines of S and Ar, since these features are easily discernible at the moderate spectral resolution of ACIS. In Table~\ref{table:lines}, we list the fluxes (corrected for absorption) obtained from the Gaussian fits; from these, we derived \Tz\ using current atomic data for a CIE plasma \citep{foster12}. In Figure~\ref{fig:overionization} (left panel), we plot the model predictions for \Tz\ as a function of line flux ratio for S and Ar, and we superimpose our measured ratios on the respective curves. For S, we find He-like to H-like ratios of 1.6--4.4, corresponding to \Tz $\approx$ 1.3--1.7 keV. We derive ratios of 2.6--9.2 for Ar; these values give \Tz $\approx$ 1.5--2.1 keV. To estimate the uncertainty in the line ratios, we refit the spectra while freezing the $kT_{\rm ej}$ to the upper and lower limits, and calculated the line ratios in these cases. In all 13 regions, the derived line ratios changed by $\sim$1\%. Therefore, we anticipate that any uncertainties are dominated by systematic errors, as discussed below. 

\begin{figure}
\includegraphics[width=\columnwidth]{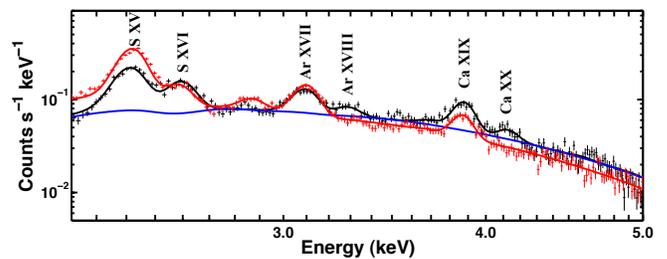}
\caption{The overionized spectrum from 2.2--5.0 keV from the top right region (region \#4) of Figure~\ref{fig:kT_e} (black points). The black line shows the best fit to this spectrum, with $kT_{\rm ISM} = 0.17$ keV, $kT_{\rm ej} = 1.20$ keV, and Gaussian components for the sulfur (with $kT_{\rm z} = 1.61$ keV) and argon ($kT_{\rm z} = 1.89$ keV). The blue line reflects the two NoLine continuum components of this fit. For comparison, we plot synthetic spectra (red points) and model of how the data would appear if the region two-temperature CIE plasmas (with $kT_{\rm ISM}= 0.17$ keV and $kT_{\rm ej} = kT_{\rm z}=$1.2 keV). In the CIE plasma spectrum, the flux in S {\sc xv} is substantially greater, and Ar {\sc xviii} is not discernible from the continuum emission.}
\label{fig:spectra}
\end{figure}

To identify regions of overionization, we map the ratio \Tz/\Tej\ across the 13 regions in Figure~\ref{fig:overionization}. Regions with \Tz/\Tej\ $>1$ are overionized, while ratios of order unity indicate ionization consistent with a CIE plasma. The degree of ionization appears to vary from east to west, with more ionized plasma toward the west of the SNR. Figure~\ref{fig:spectra} gives an example spectrum of an overionized region and a synthetic spectrum for comparison of how the data would appear if the plasma was in CIE. These two spectra have remarkable differences, including a much greater S {\sc xv} flux and much lower Ar {\sc xviii} flux in the CIE plasma spectrum than observed in the overionized spectrum. Our localization of the overionized plasma toward the west of W49B is consistent with the findings of \cite{miceli10}, who contended the central and western regions of W49B are overionized based on larger hardness ratio there (which they attribute to the presence of Fe RRCs).

To assess our ability to measure \Tej\ and \Tz, we performed the same analyses on synthetic ACIS spectra of two-temperature CIE plasmas. In particular, we produced 28 spectra over a grid of temperatures: $kT_{\rm ISM} =$ 0.25, 0.50, 0.75, and 1.00 keV and $kT_{\rm Z} =$ 1.00, 1.25, 1.50, 1.75, 2.00, 2.25, and 2.50 keV (i.e., we considered every combination of these two temperatures). To produce these 28 synthetic spectra, we assumed $N_{\rm H} = 8\times10^{22}$ cm$^{-2}$, solar abundances for the cool component, and 5$\times$ solar abundances for the hot component (values consistent with the integrated properties of W49B: \citealt{hwang00,miceli06,lopez09a}). We analyzed the 28 synthetic spectra using the same procedure as outlined above: we modeled the spectra as absorbed, two NoLine APEC components plus Gaussian functions for the emission lines. Then, we measured the S, Ar, and Ca line ratios and derived \Tej\ and \Tz\ in each spectrum, We compared the input temperatures of the synthetic spectra to our derived temperatures, and we use the deviations from CIE (where \Tej\ should equal \Tz) to evaluate the systematic errors in our measurement of \Tz. We found that for the temperatures $kT_{\rm ISM}$ and $kT_{\rm ej}$ of W49B, \Tz\ derived from the sulfur lines was within $\sim$5\% of \Tej. For argon at W49B temperatures, we obtained \Tz\ that were systematically above \Tej\ by 10--20\%, with less accuracy with lower \Tej. 

We also performed the same synthetic analyses on the He-like to H-like line flux ratios of calcium to assess whether to include these lines in our study of W49B. We found that \Tz\ derived from calcium was more than 0.5 keV greater than \Tej, at the temperatures of W49B. We note that at hotter ejecta temperatures of \Tej = 2.5 keV, \Tz\ is reliable and within 5\% of \Tej. However, given that \Tej\ of the 13 regions of W49B are $\sim$1--2 keV, we opted to exclude calcium from our analyses. 

The systematic errors partly explain why we find a larger degree of overionization in argon than in sulfur. From our synthetic spectral analyses, we anticipate that \Tz\ obtained from argon is $\sim$ 0.2 keV above the true \Tz, yet our W49B results showed \Tz\ from argon is $\sim$ 0.2--0.5 keV above that of sulfur. The origin of this difference is unlikely to be physical given that the recombination timescales of sulfur are longer than argon \cite{smithhughes10}. Although these elements yielded different magnitudes of overionization, both show the same trend of increasing ionization to the west, and the changes in ionization state we find across W49B are larger than the systematic errors. Broadly, our results indicate that the accuracy in measuring \Tz\ using particular line ratios is dependent on the plasma's electron temperature, and thus, one should consider \Tej\ when selecting which line ratios to employ to assess overionization of the plasma. We note that we have considered the case of W49B, where the cool plasma has ISM (solar) abundances. If one has two temperature plasmas and each component has super-solar abundances, the emission lines which give the most accurate measure of \Tej\ may be different than our findings above. 

\section{Discussion} \label{sec:discussion}

As noted above, two scenarios may lead to the rapid cooling associated with overionization: adiabatic expansion (e.g., \citealt{itoh89,moriya12,shimizu12}) or thermal conduction (e.g., \citealt{cox99,shelton99}). The maps of ionization state across W49B in Figure~\ref{fig:overionization} can aid in distinguishing which of these mechanisms may dominate. In particular, if the cooling originates from adiabatic expansion, overionization should be most prevalent where the blast wave is expanding through a rarefied ISM. Conversely, if the plasma cools via thermal conduction, the overionized hot gas should be concentrated where it is in contact with cold, dense material. 

\begin{figure*}
\includegraphics[width=\textwidth]{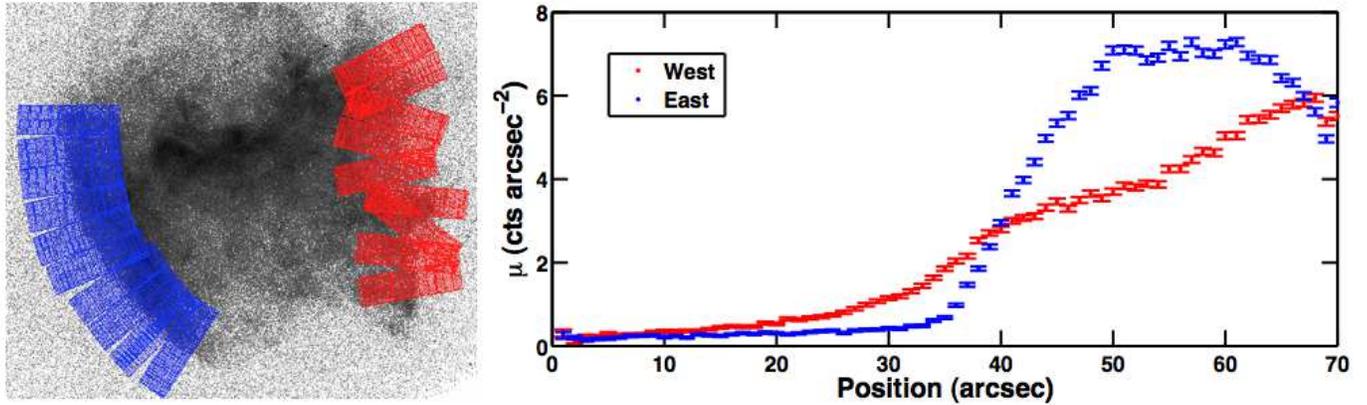}
\caption{Left: Regions used to analyze the surface brightness profiles for the east (blue) and west (red) of W49B. Right: Mean surface brightness profiles of the full band (0.5--8.0 keV) across the east (blue) and west (red) as a function of position. The position of 0\arcsec\ is defined as the outer edge of the regions.}
\label{fig:profiles}
\end{figure*}

We find a gradient of increasing overionization toward the west of W49B, while the east is at or near CIE. Given the location of molecular material and the plasma temperatures across W49B, the distribution of the overionized plasma is most consistent with the adiabatic cooling scenario. In particular, the ejecta are impacting nearby molecular material (see the blue contours in Figure~\ref{fig:kT_e}, left) in the east, and the plasma is the hottest there (with  \hbox{\Tej $\sim$1.8 keV}: Figure~\ref{fig:kT_e}, right). In the west, the ejecta are expanding relatively unimpeded (particularly in the northwest), and the plasma is the coolest there (with \hbox{\Tej $\ls$1.2 keV}). Although molecular material is seen in the southwest, this gas may be seen in projection: low-frequency radio observations have shown intervening absorbing material there \citep{lacey01} and the $N_{\rm H}$ is enhanced in that area of the SNR (with $N_{\rm H} \gs 8 \times10^{22}$ cm$^{-2}$: \citealt{lopez12}). Our results are consistent with the recent hydrodynamical simulations performed by \cite{zhou11}, who showed that the spatial distribution of overionized ejecta can be attributed to the rapid free expansion of the plasma following heating by an inhomogeneous medium (although their models focus principally on the central region of the SNR).  

We are unable to confirm the expansion rates of the two sides of W49B using proper motion: the 11-year baseline between {\it Chandra} observations is not long enough to observe expansion with the spatial resolution of ACIS. At the off-axis position of the SNR rim, the ACIS point-spread function is $\sim$4 pixels\footnote{The radius where 90\% of encircled energy of a source is detected at 1.49 keV: http://cxc.harvard.edu/proposer/POG/html/index.html}. Thus, a velocity of $\gs$6800 km s$^{-1}$ would be necessary to detect the expansion, assuming a distance $D=$ 8 kpc to W49B \citep{moffett94,hwang00}. 

However, the surface brightness profiles across the SNR boundaries are consistent with impeded expansion in the east and faster expansion in the west (see Figure~\ref{fig:profiles}). For this analysis, we computed the net full-band surface brightness $\mu$ (in counts arcsec$^{-2}$) across nineteen profiles, each composed of 68 narrow rectangles with area 2$\times$40 pixel$^{2}$ (Figure~\ref{fig:profiles}, left). We oriented the rectangles such that they were parallel to the X-ray full-band contours, and we averaged the profiles of the east and the west (after translating the maxima of the profiles so they were aligned). Figure~\ref{fig:profiles} (right) gives the resulting mean $\mu$ value as a function of position across the east and west edge of W49B. The eastern profile has a steep slope, with a factor of $\sim$10 increase in $\mu$ over $\sim$15\arcsec. By contrast, the change in $\mu$ in the west is much more gradual, with a factor of $\sim$3 increase in $\mu$ over $\sim$35\arcsec. The sharp boundary in the east is consistent with the ejecta being inhibited by the molecular material there, whereas the emission in the west is more diffuse. 

\section{Summary} \label{sec:summary}

We have presented maps of the overionized plasma in W49B using data from the recent 220 ks {\it Chandra} ACIS observation. By comparing the temperatures derived from the bremsstrahlung continuum to those from the He-like to H-like line ratios of sulfur and argon, we found that the SNR plasma is overionized in the west and has a gradient of increasing ionization from east to west. Based on the impeded expansion of the ejecta in the east and relatively lower inferred density ISM in the west, we suggest that the overionization maps are consistent with adiabatic expansion being the dominant cooling mechanism. In the future, these results can be confirmed by performing spatially-resolved spectroscopic analysis of the Fe RRC in a deeper {\it XMM-Newton} observation of W49B and by measuring directly the expansion velocities of the SNR through proper motion or spectroscopy studies.

\acknowledgements

Support for this work was provided by National Aeronautics and Space Administration through Chandra Award Number GO2--13003A and through Smithsonian Astrophysical Observatory contract SV3--73016 to MIT issued by the Chandra X-ray Observatory Center, which is operated by the SAO for and on behalf of NASA under contract NAS8--03060. We acknowledge support from the David and Lucile Packard Foundation and NSF grant AST--0847563. Support for LAL was provided by NASA through the Einstein Fellowship Program, grant PF1--120085, and the MIT Pappalardo Fellowship in Physics.

\end{document}